\documentclass[12pt,manuscript]{aastex}
\shorttitle{Radio Pulsation Search and Imaging Study of SGR J1935+2154}
\shortauthors{Surnis et al.}
\usepackage{natbib}
\usepackage{fixltx2e}
\newcommand{\sups}[1]{\textsuperscript{#1}}

\begin{document}

\title{Radio Pulsation Search and Imaging Study of SGR J1935+2154}
\author{Mayuresh. P. Surnis}
\affil{National Centre for Radio Astrophysics, Tata Institute of Fundamental Research, P. O. Bag 3, University of Pune Campus, Ganeshkhind, Pune, India.}

\author{Bhal Chandra Joshi}
\affil{National Centre for Radio Astrophysics, Tata Institute of Fundamental Research, P. O. Bag 3, University of Pune Campus, Ganeshkhind, Pune, India.}

\author{Yogesh Maan\altaffilmark{1}}
\affil{National Centre for Radio Astrophysics, Tata Institute of Fundamental Research, P. O. Bag 3, University of Pune Campus, Ganeshkhind, Pune, India.}

\author{M. A. Krishnakumar}
\affil{Radio Astronomy Centre, National Centre for Radio Astrophysics, Tata Institute of Fundamental Research, Ootacamund, India.}

\author{P. K. Manoharan}
\affil{Radio Astronomy Centre, National Centre for Radio Astrophysics, Tata Institute of Fundamental Research, Ootacamund, India.}

\and

\author{Arun Naidu}
\affil{National Centre for Radio Astrophysics, Tata Institute of Fundamental Research, P. O. Bag 3, University of Pune Campus, Ganeshkhind, Pune, India.}

\altaffiltext{1}{ASTRON, Netherlands Institute for Radio Astronomy, Postbus 2, 7990 AA Dwingeloo, The Netherlands}

\begin{abstract}

We present the results obtained from imaging observations, and search for radio pulsations towards the magnetar SGR J1935+2154 made using the Giant Metrewave Radio Telescope, and the Ooty Radio Telescope. We present the high resolution radio image of the supernova remnant (SNR) G57.2+0.8, which is positionally associated with SGR J1935+2154. We did not detect significant periodic radio pulsations from the magnetar, with 8$\sigma$ upper limits on its flux density of 0.4, and 0.2 mJy at 326.5, and 610 MHz, respectively, for an assumed duty cycle of 10\%. The corresponding 6$\sigma$ upper limits at the two frequencies for any burst emission with an assumed width of 10 ms are 0.5 Jy, and 63 mJy, respectively. No continuum radio point source was detected at the position of SGR J1935+2154 with a 3$\sigma$ upper limit of 1.2 mJy. We also did not detect significant diffuse radio emission in a radius of 70 arc seconds in coincidence with the diffuse X-ray emission reported recently, with a 3$\sigma$ upper limit of 4.5 mJy. Using the archival HI spectra, we estimate the distance of SNR G57.2+0.8 to be 11.7 $\pm$ 2.8 kpc. Based on measured HI column density (N$_H$) along this line of sight, we argue that the magnetar could be physically associated with SNR G57.2+0.8. Based on present data, we can not rule out either a pulsar wind nebula or a dust scattering halo origin for the diffuse X-ray emission seen around the magnetar.

\end{abstract}


\section{Introduction}
\label{intro}
Soft gamma-ray repeaters (SGRs) are sources that produce repeating bursts of $\gamma$-rays and are usually discovered as $\gamma$-ray transients. Together with anomalous X-ray pulsars (AXPs), they form a rare, distinct class of pulsars (with pulsations in a narrow period range of 2$-$12 s, and $\dot{P}$ between 10\sups{-14}$-$10\sups{-11}), which have transient as well as highly variable emission in the high energy electromagnetic band. \cite{dt92} proposed that the object powering the high energy bursts could be a neutron star with the magnetic field strength in the range 10\sups{14}$-$10\sups{15} G (hence the adopted name of magnetar), and the bursts are powered by the release of magnetic energy in a short time scale. The burst luminosity of such a source is typically higher than the spin-down luminosity inferred from the measurement of its period, and period-derivative.  

Most magnetars are radio-quiet. First detection of radio pulsations from a magnetar was from XTE J1810$-$197, following a 100-fold increase in its X-ray flux in 2003 \citep{ims+04,crh+06}. There are now three such sources with reported radio pulsations, which also show a large variability in their emission, spectral indices as well as integrated profile shapes at radio frequencies \citep{ssw+09,ccr+07,crh+16}. In all these cases, the quiescent X-ray luminosity is less than the spin-down luminosity \citep{rpt+12}. Radio emission properties similar to the ones above are also observed in PSR J1622$-$4950. This high magnetic field pulsar is the first object of this class, which was discovered from its radio emission \citep{lbb+10,lbb+12}. Thus, all the known radio loud magnetars have shown X-ray outbursts followed by pulsed radio emission. The above observational evidence suggests that the onset of X-ray bursts in a magnetar  qualifies it as a plausible candidate for radio follow-up observations.

Generally, a young pulsar powered by its rotation, is enveloped by a bright synchrotron emitting nebula, created by the wind of relativistic particles from the pulsar, injected, and powered by the loss of rotational kinetic energy of the pulsar \citep{gol69, ps73}. Even though the spin-down luminosity of magnetars is smaller than young pulsars, a wind nebula could still be present around them. There exists no unambiguous detection of such a wind nebula so far \citep[for more details, see][ and references therein]{ykk+12}. A magnetar burst is also likely to create a transient nebula just after the burst \citep{fkb99}. Recently, a candidate nebula has been proposed for SWIFT J1834$-$0846 \citep{ykk+12}, which could also be explained as a halo created by the scattering of X-rays by intervening dust \citep{etr+13}. Thus, imaging observations in a wide band of radio frequencies following an SGR burst can be useful for detecting a magnetar powered wind nebula, and constraining the nature of diffuse X-ray emission. 

The Swift burst alert telescope (BAT) detected the X-ray burst of SGR J1935+2154 on 2014 July 5. The duration of the burst was $\sim$0.1 s with a double peaked structure \citep{cbc+14}. Following the trigger given by the above BAT detection, the Swift X-ray telescope (XRT) also observed the field. The XRT, and ultraviolet and optical telescope (UVOT) imaging observations were useful to localize the source at the right ascension (RA): 19\sups{h}34\sups{m}55\sups{s}.68, and declination (DEC): +21$^{\circ}$53'48".2. The initial burst was also followed by three short interval soft bursts, out of which the first two had durations of 30 $\mu$s each, while the third lasted for 70 $\mu$s. Additionally, \cite{cc14} analyzed the BAT archival data and found no detection. However, they could find serendipitous detection of the same source in Swift XRT data made in years 2010, and 2011. It was also noted that the SGR lies very close to the geometric center of the supernova remnant (SNR) G57.2+0.8, and there is a good chance of association. The age of this SNR is undetermined and its distance is estimated to be 9 $\pm$ 4 kpc using $\sigma-\Delta$ relation \citep{srr+11,puv+13}. Recently, based on {\it Chandra}, and {\it XMM-Newton} data, \cite{ier+16} reported coherent pulsations from this source with a period of 3.2 s. They confirmed its magnetar nature with a slow-down rate of 1.43 $\times$ 10$^{-11}$ s s$^{-1}$, and a spin-down luminosity, L$_{sd} \sim $ 1.7 $\times$ 10$^{34}$ erg s$^{-1}$. They also reported diffuse X-ray emission extending upto 1' around the magnetar, which could be a dust scattered halo or pulsar wind nebula (PWN).

In this paper, we report on the results of the radio pulsation search, and radio imaging observations carried out using the Giant Metrewave Radio Telescope (GMRT), and the Ooty Radio Telescope (ORT). Details of the observations are presented in Section \ref{obs}. Details of the data analysis procedure, results obtained, and the implications of non-detection of diffuse radio emission are discussed in Section \ref{rnd}, followed by a summary of our findings in Section \ref{sum}.

\section{Observations}
\label{obs}
We performed observations of SGR J1935+2154 with the ORT at 326.5 MHz on 2014 July 9, and 14, using 16 MHz bandwidth in the pulsar mode. The data were recorded using the PONDER \citep{njm+15} back-end, employing 1024 filterbank channels with a sampling time of 1 ms. The total integration time in the two sessions was 4, and 3 hours, respectively. Observations with the GMRT were performed using the director's discretionary time (DDT) allocation (ddtB134). The observations were performed on 2014 July 14, at 610 MHz with 33 MHz band spread across 512 filterbank channels using the GMRT software back-end \citep{rgp+10}. The interferometric data were recorded with a default sampling of 16 s, while the pulsar data were recorded in the phased array (PA), and incoherent array (IA) modes with a sampling time of 61 $\mu$s. The high sensitivity PA beam (100") was centred at the initial SWIFT position (RA: 19\sups{h}34\sups{m}55\sups{s}.68, DEC: +21$^{\circ}$53'48".2), whereas the IA beam covered the entire 40' field. The total integration time was 2 hours, out of which on-source time was 1.25 hours. In the imaging mode, 3C48, and 1822$-$096 were observed as flux density, and phase calibrators, respectively. PSR B1937+21 was observed as a control pulsar to estimate the sensitivity of the pulsar-mode data.

\section{Results \& Discussion}
\label{rnd}

\subsection{Data Analysis}
\label{anal}
The pulsar data obtained with both the GMRT and the ORT were analysed using the analysis pipeline developed for the NCRA high performance cluster (HPC) based on SIGPROC\footnote{http://sigproc.sourceforge.net} pulsar data analysis package. In order to fully exploit the high time resolution of the data, the dispersion measure (DM) step used for the pulsar search had to be very fine. The data were thus, dedispersed to 2240 trial DMs in the range 0$-$1500 pc cm$^{-3}$. Such a DM limit was imposed keeping in mind the very high DM measured for the Galactic centre magnetar, PSR J1745$-$2900 \citep{efk+13} and the fact that patchy nature of ionised plasma structures along the line of sight may produce a DM which may be much more than the model prediction. The dedispersed time-series data were then searched using both harmonic, and single pulse search. The resultant candidate plots were manually scrutinised to identify the presence of pulsar signatures.

\begin{figure}[h]
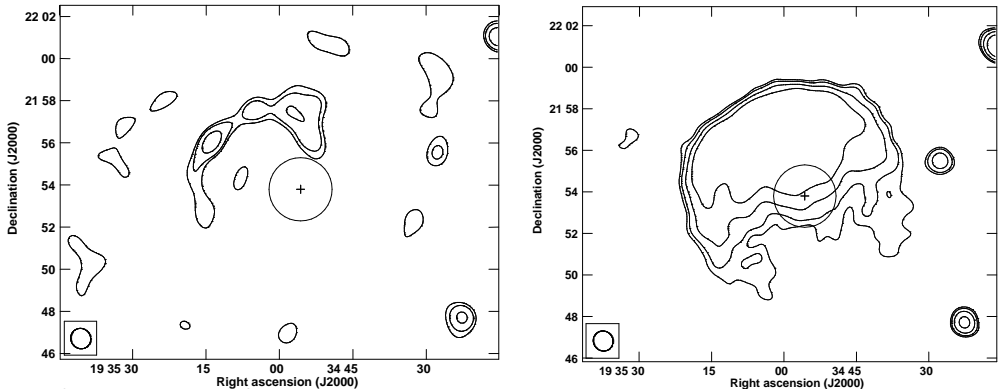

\begin{center}

\includegraphics[trim=0 1.2cm 0 1.2cm, clip, scale=0.36]{G57_gmrt.eps}
\includegraphics[trim=0 1.25cm 0 1.2cm, clip, scale=0.34]{G57_vla.eps}
\caption{Left: GMRT Image of SNR G57.2+0.8 at 610 MHz, convolved with a circular Gaussian with an FWHM of 60". The RMS noise in the map is 4 mJy. The contour levels are at 3, 5, 10, and 20$\sigma$. Right: VLA Image of SNR G57.2+0.8 at 1365 MHz, obtained in the D configuration (see text for details). The RMS noise in the map is 0.6 mJy. The contour levels are at 3, 5, 10, and 20$\sigma$. In both images, the cross indicates the position of SGR J1935+2154 with the size of the cross indicating 10$\sigma$ position uncertainty, and the solid circle represents the extent of diffuse X-ray emission reported by \cite{ier+16}.}
\label{lowres}

\end{center}
\end{figure}

The imaging data were flagged, and calibrated using FLAGCAL\footnote{http://ncralib1.ncra.tifr.res.in:8080/jspui/handle/2301/581} \citep[e.g.][]{c13}, and the image was made using the multi-facet imaging technique in the Astronomical Image Processing System (AIPS)\footnote{http://www.aips.nrao.edu/index.shtml} with 19 facets spread across the primary beam area of about 0.3 square degree. The CLEAN image obtained after the de-convolution was further improved by self calibration. The time scales for calculating the phase solutions were 5, 1, and 0.5 minutes respectively for the three iterations of self calibration. The resultant high resolution image is shown in Figure \ref{hires}. We made another image of SNR G57.2+0.8 from the archival data sets\footnote{https://science.nrao.edu/facilities/vla/archive/index} obtained from the Very Large Array (VLA) at 1365 MHz. The VLA observations were made on 1995 April 6 over a bandwidth of 25 MHz, using the VLA in the D configuration, with a maximum baseline of 5 k$\lambda$, and an angular resolution of $\sim$60" (Project code: AH0535, PI: Mark Holdaway). In order to compare with the VLA image, the GMRT image was convolved with a circular Gaussian beam of a full width at half maximum (FWHM) of 60". This was achieved using the CONVL\footnote{http://www.aips.nrao.edu/cgi-bin/ZXHLP2.PL?CONVL} routine of AIPS. The resultant images having equal resolution are shown in Figure \ref{lowres}.

Additionally, publicly available archival spectral cubes were downloaded from the VLA Galactic Plane Survey \citep[VGPS;][]{std+06} to obtain HI spectra towards the SNR. Spectra from the bright northern portion of SNR as well as a blank region just above it were extracted at 5 pixels each (marked by boxes in Figure \ref{radioX}), to form the ON, and OFF spectra. These were then used to obtain the emission, and absorption spectra towards the source using a procedure similar to that described in \cite{csd+13}. The average emission spectrum obtained towards the direction of SGR J1935+2154 was used to calculate the HI column density (N$_H$) along this line of sight. This was done as follows. First, the moment zero map was generated from the spectral cube. The RMS was determined in this map over an area roughly equal to that of the diffuse emission seen in X-ray image (which was found to be 2.7 K). N$_H$ was then calculated assuming an optically thin medium, implying \citep{vkv74}

\begin{equation}
N_H \mbox{ } (cm^{-2}) = 1.823 \times 10^{18} \int T_{B} dV
\end{equation} 
where T$_B$ is the brightness temperature in Kelvin, and V is the radial velocity in km s$^{-1}$. In the emission profile obtained towards the SGR, all the bins having brightness temperature more than 2.7 K were summed together and then multiplied by the total range of velocity bins to get an estimate of N$_H$ (total area under the emission profiles).

The public archival data \citep[obtained by][]{ier+16} from the PN charge coupled device (CCD) instrument \citep{sbd+01} of the European Photon Imaging Camera (EPIC) on board the {\it XMM-Newton} mission for all but one observation (as the latest observational data were not public) were added together to make a 98 ks exposure image of the region around SGR J1935+2154. This image is shown in Figure \ref{radioX}. More details about the {\it XMM-Newton} observations are given in Table 1 of \cite{ier+16}. The diffuse emission around SGR J1935+2154, which was reported by \cite{ier+16}, is seen clearly in this image. 

\subsection{Results}
\label{res}
The time-series observations, both from the ORT and the GMRT, did not show the presence of significant pulsations. The flux density upper limits for a putative radio pulsar (associated or otherwise) were then calculated for both GMRT (IA and PA) and ORT data using the radiometer formula. The flux density limits thus derived, are applicable to the highest DM searched for normal pulsars at both frequencies. With an assumed 10\% duty cycle, these limits for the high resolution GMRT data are applicable to millisecond pulsars only upto a DM of 100 pc cm$^{-3}$. For periodic pulsed emission, the 8$\sigma$ upper limits on average flux density assuming 10\% duty cycle are 0.4, and 0.2 mJy at 326.5, and 610 MHz, respectively. The GMRT PA data were subsequently folded at the period detected by \cite{irz+14} and it also resulted in a non-detection. The preliminary results for the pulsation search were reported earlier in the Astronomer's Telegram \citep{skm+14}. We also did not detect significant bursts with 6$\sigma$ upper limits on flux density (assuming 10 ms burst duration) of 0.5 Jy, and 63 mJy at 326.5, and 610 MHz, respectively. The corresponding flux density upper limits from GMRT IA data at 610 MHz are 0.8 mJy (8$\sigma$) for pulsed emission (with an assumed duty cycle of 10\%), and 244 mJy (6$\sigma$) for isolated bursts (assuming 10 ms burst duration).\\

\begin{figure}[h]
\begin{center}

\includegraphics[trim=0 1.2cm 0 1.2cm, clip, scale=0.7]{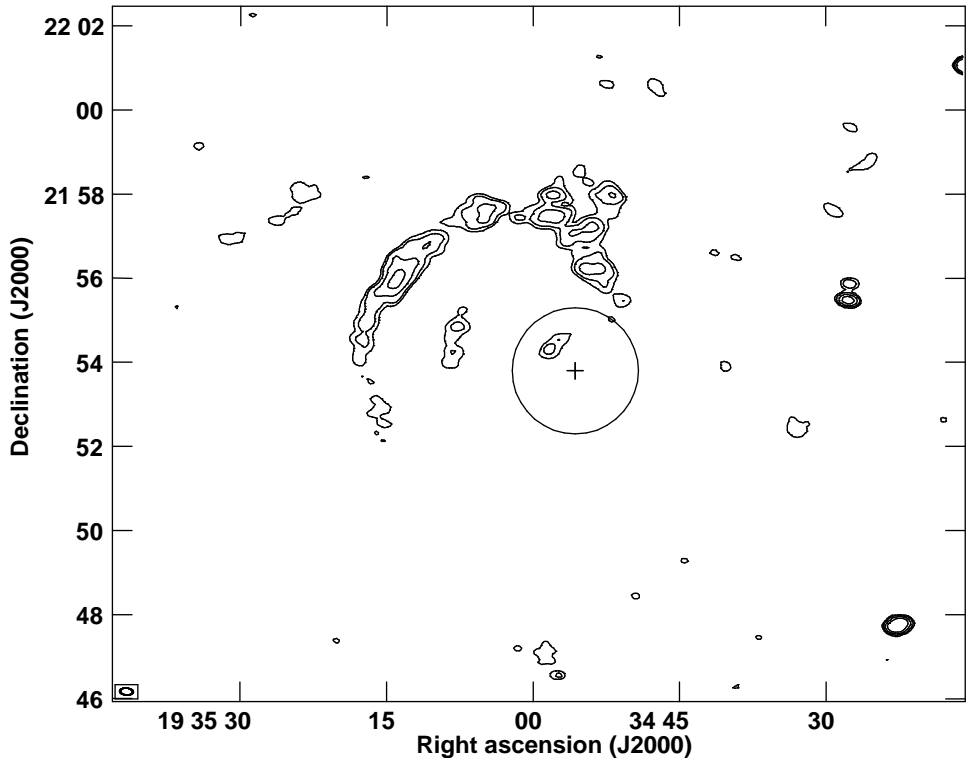}
\caption{High resolution GMRT Image of SNR G57.2+0.8 at 610 MHz. The RMS noise in the map is 0.4 mJy. The contour levels are at 3, 5, 10, and 20$\sigma$. The cross indicates the position of SGR J1935+2154 with the size of the cross indicating 10$\sigma$ position uncertainty. The solid circle represents the extent of diffuse X-ray emission reported by \cite{ier+16}.}
\label{hires}

\end{center}
\end{figure}

\begin{figure}[h]
\begin{center}

\includegraphics[scale=0.6]{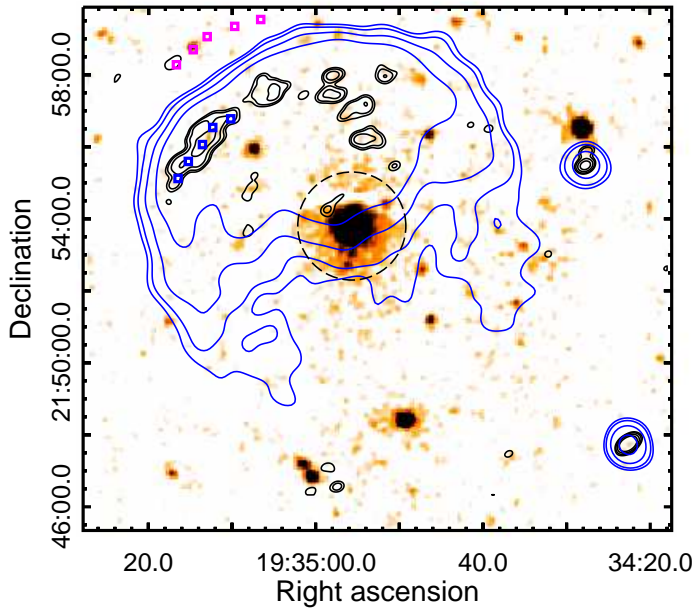}
\caption{98 ks integrated {\it XMM-Newton} EPIC-PN image of the region around SGR J1935+2154. The publicly available archival data were obtained by G. L. Israel \citep{ier+16} and the observation details are given in Table 1 therein. The resultant map has been smoothed using a Gaussian function with a radius of 4". The blue contours are plotted from the 1365 MHz VLA map of SNR G57.2+0.8 (as shown in the right panel of Figure \ref{lowres}), while black contours are plotted from the high resolution GMRT image at 610 MHz (as shown in Figure \ref{hires}). The pixels used for obtaining the ON, and OFF spectra are marked by blue, and magenta boxes, respectively. The dashed circle represents a region of radius 90", which includes the diffuse X-ray emission (see text for more details).}
\label{radioX}

\end{center}
\end{figure}

We have made the first high resolution image of SNR G57.2+0.8 at 610 MHz (Figure \ref{hires}). The synthesized beam size is 18.5" $\times$ 9.9" with a parallactic angle of 85\sups{$\circ$} (as indicated in the bottom left corner of the image in Figure \ref{hires}), while the root mean square (RMS) noise in the map is 0.4 mJy. As it can be clearly seen, there is no associated radio point source at the position of SGR J1935+2154, with a 3$\sigma$ upper limit of 1.2 mJy. Thus, the non-detection inferred from the time-series is consistent with the absence of a continuum source in the radio map. No diffuse radio emission, associated with a putative wind nebula with an extent similar to the diffuse X-ray emission reported by \cite{ier+16} (in radii of 15", and 70"), was detected with a 3$\sigma$ flux density upper limit of about 4.5 mJy over a circular area with a radius of 70".

\begin{figure}[h]
\begin{center}

\includegraphics[trim=0 0cm 0 0cm, clip, scale=0.45]{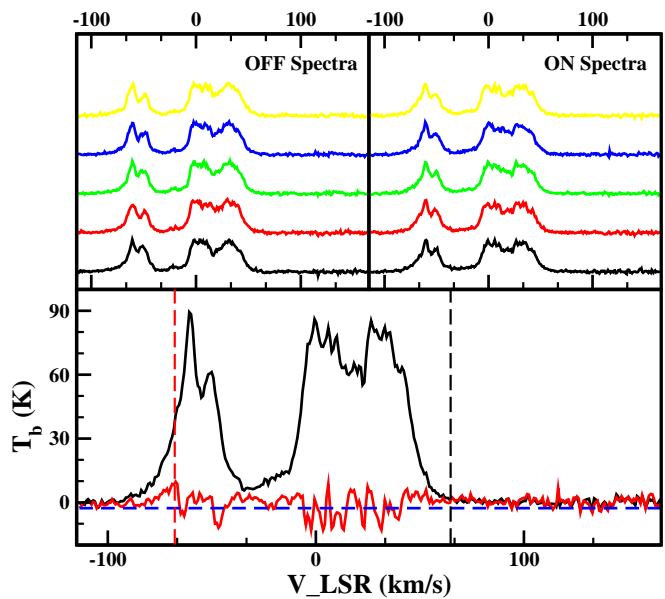}
\caption {Emission spectra, and absorption profile extracted from VGPS archival data \citep{std+06}. The top two panels show the extracted spectra from 5 pixels each for the OFF (left panel), and the ON source region (right panel; as indicated in Figure \ref{radioX}). The bottom panel shows the averaged ON source emission spectrum (black), and the absorption profile (red) towards the northern filament of SNR G57.2+0.8. The black dashed line shows the maximum velocity at the tangent point, the dashed red line indicates the absorption feature with the highest negative velocity, and the dashed blue line indicates $-$2.7 K below which, absorption features are believable.}
\label{HIprof}

\end{center}
\end{figure}

The HI emission spectra along with the absorption profiles are shown in Figure \ref{HIprof}. The average spectrum shows a broad emission feature upto a velocity of 61.2 km s$^{-1}$, which is the maximum velocity at the tangent point. As there are clear absorption features present at negative radial velocities, the continuum source (the SNR in this case) lies outside the solar orbit, much beyond the tangent point. The absorption features appear upto the maximum negative velocity of -63.3 km s$^{-1}$. This translates to a kinematic distance of 11.7 $\pm$ 2.8 kpc (1$\sigma$ uncertainty), using the circular rotation curve of the Milky Way given by \cite{fbs89} (assuming $\Theta$ = 220 km s$^{-1}$, R$_{\Sun}$ = 8.5 kpc). 

From HI emission profile obtained in the direction of SGR J1935+2154, we calculate the HI column density in this line of sight to be (1.1 $\pm$ 0.1) $\times$ 10$^{22}$ cm$^{-2}$.

\subsection{Discussion}
\label{disc}
Our simultaneous imaging, and time-series observations with the GMRT and the ORT resulted in non-detection of radio pulsations or a radio continuum source in the direction of SGR J1935+2154. During 4 to 10 days after the X-ray burst was reported in this source, the upper limits on average flux density for pulsed emission, assuming 10\% duty cycle, are 0.4, and 0.2 mJy at 326.5, and 610 MHz, respectively. The upper limit on the flux density of associated continuum point source is 1.2 mJy at 610 MHz. Moreover, no extended radio emission is discernible in the 610 MHz map either within a radius of 15" or 70" with an upper limit of flux density over a 70" radius of 4.5 mJy. We also estimate the distance to the SNR to be 11.7 $\pm$ 2.8 kpc. This distance is consistent with the distance estimated by \cite{puv+13}.

The SGR J1935+2154 lies close to the centroid of SNR G57.2+0.8. An approximate estimate of distance to the magnetar can be made as follows. The spectral analysis by \cite{ier+16} obtained a value of N$_H$ as 2 $\times$ 10$^{22}$, and 1.6 $\times$ 10$^{22}$ cm$^{-2}$ from their spectral fits using {\it Chandra} and {\it XMM-Newton} data, respectively towards the magnetar, and 3.8 $\times$ 10$^{22}$ cm$^{-2}$ for the diffuse emission from {\it XMM-Newton} data. If we assume that the N$_H$-DM correlation \citep{hnk13} holds for this line of sight, we can obtain an estimate of associated DM for the magnetar. Adopting the value of N$_H$ as 1.8 $\times$ 10$^{22}$ cm$^{-2}$, we get a DM of 600 $^{+260}_{-180}$ pc cm$^{-3}$ for the magnetar, and 1300 $^{+500}_{-400}$ pc cm$^{-3}$ for the diffuse emission. Using NE2001 model \citep{cl02} towards this line of sight, we get a distance estimate for both the magnetar, and the diffuse emission to be much higher than the extent of the Galaxy ($>$50 kpc). However, it is to be noted that the NE2001 model saturates beyond a DM of 540 pc cm$^{-3}$ in this line of sight. Also the distance estimates from the N$_H$-DM correlation can have a large uncertainty beyond a DM of 100 pc cm$^{-3}$ or a distance greater than 3 kpc \citep{hnk13}. \cite{hnk13} also note that the N$_H$ value inferred from X-ray observations of pulsars at low Galactic latitudes may be higher than that inferred from HI emission profiles. This happens due to absorption of X-rays by molecular clouds rather than neutral Hydrogen. With this caveat, the N$_H$ values measured by \cite{ier+16} for the magnetar, and our own measurements along the same line of sight are consistent with each other. Considering the spiral structure of the HI in our Galaxy \citep{lbh06}, and the N$_{H}$ values obtained from the Leiden Argentine Bonn (LAB) survey\footnote{https://www.astro.uni-bonn.de/hisurvey/profile/} \citep{kbh+05} along different lines of sight in the Galactic latitude range of 40$^{\circ}-$80$^{\circ}$, which are all nearly around 1$-$1.5 $\times$ 10$^{22}$ cm$^{-2}$, the magnetar (and the diffuse emission) could be located in or beyond the Perseus arm. With all these (uncertain) estimates of the magnetar distance, it can be argued that SGR J1935+2154 and SNR G57.2+0.8 may be physically associated.

\cite{ier+16} estimated the quiescent X-ray luminosity of the magnetar, L$_X$ as 5 $\times$ 10$^{33}$ ergs s$^{-1}$ assuming a distance of 9 kpc. This distance was derived assuming an association between the magnetar and the SNR. Even with our revised estimate for the SNR distance, such an association is possible. As discussed by \cite{ier+16} assuming a distance of 9 kpc, the SGR is somewhat under-luminous during a burst as compared to a typical magnetar. It may be noted that this luminosity is still larger than 10 out of 22 magnetars listed in online McGill magnetar catalog\footnote{http://www.physics.mcgill.ca/$\sim$pulsar/magnetar/main.html} \citep{ok14}. The SGR was observed from Parkes telescope at 10, and 20 cm wavelength by \cite{bir+14}, resulting in a non-detection with 8$\sigma$ upper limits of 0.07, and 0.1 mJy at 10, and 20 cm, respectively, which is consistent with our non-detection at 610, and 326.5 MHz. With a typical flat spectrum emission expected from radio-loud magnetars \citep[e.g.][]{ydw+15}, our upper limits for the pulsed radio emission along with the high frequency limits may imply a spectral index less steeper than $-$0.8. Assuming an association between the SGR and the SNR, and considering the upper limit on the distance of the magnetar to be 14.5 kpc (from 1$\sigma$ error on the SNR distance), the X-ray luminosity of the SGR turns out to be 1.3 $\times$ 10$^{34}$ ergs s$^{-1}$. Even with a slightly increased estimate, the ratio L$_{X,qui}$/L$_{sd}$ is still about 0.65, which is much less than unity. Thus, the SGR falls into the ``radio-loud'' class and the non-detection of radio pulsations are puzzling, but could
be due to the radio beam pointing away from the line-of-sight.

No radio counterpart of diffuse X-ray emission reported by \cite{ier+16} was detected in the high resolution image at 610 MHz. The diffuse X-ray emission could be due to the scattering of X-rays from the magnetar by a dust halo \citep{etr+13,ier+16} or a wind nebula powered by the magnetar \citep{ykk+12,ier+16}. In the former case, one does not expect diffuse radio emission, which is consistent with our non-detection. In case of a wind nebula, synchrotron radio emission is expected. As a PWN is powered by the loss of rotational kinetic energy from the central pulsar, one can expect its radio luminosity to be related to its spin-down luminosity. Assuming this proportionality to be in the form S = K $\times$ $\dot{E}$ / d$^2$ for an order of magnitude calculation, where S is the flux density in Jy at 1 GHz, $\dot{E}$ is the spin-down luminosity in ergs/s and d the distance to the PWN in kpc, one can estimate the value of proportionality constant K for some typical PWNs such as Crab, Vela and Kes75. With the spin-down luminosities from Australia Telescope National Facility (ATNF) pulsar catalog\footnote{http://www.atnf.csiro.au/people/pulsar/psrcat/}\citep{mht+05} and distances and 1 GHz flux density from available literature \citep{cgk+01,dlr+03,gre14,lt08,mil68}, we obtain an average value of K as 1.6 $\times$ 10$^{-15}$. With our distance estimate of 11.7 kpc and spin-down luminosity of SGR J1935+2154 of 2 $\times$ 10$^{34}$ ergs s$^{-1}$ \citep{ier+16}, we expect a flux density of 2.4 mJy at 610 MHz (assuming a spectral index of $-$0.3) for any associated PWN. As the upper limit on the flux density of diffuse emission from our observations is higher than this, we can not rule out the presence of such an emission. We would like to caution that this conclusion is based on the assumption that the radio luminosity is proportional to spin-down luminosity. Such a relation is uncertain as it would depend upon many factors, such as integrated spin-down history of pulsar, nature of confining reverse shock and pulsar velocity. The nature of a wind nebula around a magnetar is also likely to be different from usual PWNs. In the light of the uncertainty in the distance estimate of the SGR, and the diffuse X-ray emission, as well as the lower expected radio PWN flux density than our upper limits, it is difficult for us to comment on the plausibility of a PWN, and we can not rule out any of these two scenarios based on the currently available data. 

\section{Summary}
\label{sum}
In summary, we have observed SGR J1935+2154 with the GMRT and the ORT at 610, and 326.5 MHz, respectively. Using these observations, we have put upper limits of 0.4, and 0.2 mJy on periodic radio pulsations from the SGR at 326.5, and 610 MHz, respectively. We also did not detect any significant burst or transient emission with 6$\sigma$ upper limits on the flux density for an assumed width of 10 ms to be 0.5 Jy, and 63 mJy at 326.5, and 610 MHz, respectively. From high resolution radio images, we put 3$\sigma$ upper limits of 1.2, and 4.5 mJy on the continuum radio flux density of the SGR, and that associated with the diffuse X-ray emission, respectively. Using HI emission, and absorption spectra, we have determined the distance of SNR G57.2+0.8 to be 11.7 $\pm$ 2.8 kpc. Based on the observed N$_H$ values for the magnetar, we argue that the magnetar could be physically associated with the SNR.

\section{Acknowledgements}
\label{ack}
We thank the staff of the ORT, and the GMRT who have made these observations possible. The GMRT is run by the National Centre for Radio Astrophysics of the Tata Institute of Fundamental Research. We thank the anonymous referee, whose suggestions have improved the manuscript. MPS would like to thank Ishwara Chandra, and Ruta Kale for the help regarding advanced imaging techniques. This work made use of PONDER receiver, which was funded by XII plan grant TIFR 12P0714. This work made use of the NCRA high performance cluster (HPC) facility, which was funded by XII plan grant TIFR 12P0711. We would like to thank V. Venkatasubramani, Shekhar Bachal and the whole HPC team. BCJ and PKM acknowledge support from DST-SERB Extra-mural grant EMR/2015/000515. For a part of the research, YM acknowledges use of the funding from the European Research Council under the European Union's Seventh Framework Programme (FP/2007-2013) / ERC Grant Agreement no. 617199.
\bibliography{ref}

\begin{thebibliography}{}
\expandafter\ifx\csname natexlab\endcsname\relax\def\natexlab#1{#1}\fi

\bibitem[{{Burgay} {et~al.}(2014){Burgay}, {Israel}, {Rea}, {Possenti},
  {Zelati}, {Esposito}, {Mereghetti}, \& {Tiengo}}]{bir+14}
{Burgay}, M., {Israel}, G.~L., {Rea}, N., {et~al.} 2014, The Astronomer's
  Telegram, 6371, 1

\bibitem[{{Camilo} {et~al.}(2006){Camilo}, {Ransom}, {Halpern}, {Reynolds},
  {Helfand}, {Zimmerman}, \& {Sarkissian}}]{crh+06}
{Camilo}, F., {Ransom}, S.~M., {Halpern}, J.~P., {et~al.} 2006, \nat, 442, 892

\bibitem[{{Camilo} {et~al.}(2007){Camilo}, {Cognard}, {Ransom}, {Halpern},
  {Reynolds}, {Zimmerman}, {Gotthelf}, {Helfand}, {Demorest}, {Theureau}, \&
  {Backer}}]{ccr+07}
{Camilo}, F., {Cognard}, I., {Ransom}, S.~M., {et~al.} 2007, \apj, 663, 497

\bibitem[{{Camilo} {et~al.}(2016){Camilo}, {Ransom}, {Halpern}, {Alford},
  {Cognard}, {Reynolds}, {Johnston}, {Sarkissian}, \& {van Straten}}]{crh+16}
{Camilo}, F., {Ransom}, S.~M., {Halpern}, J.~P., {et~al.} 2016, \apj, 820, 110

\bibitem[{{Castelletti} {et~al.}(2013){Castelletti}, {Supan}, {Dubner},
  {Joshi}, \& {Surnis}}]{csd+13}
{Castelletti}, G., {Supan}, L., {Dubner}, G., {Joshi}, B.~C., \& {Surnis},
  M.~P. 2013, \aap, 557, L15

\bibitem[{Chengalur(2013)}]{c13}
Chengalur, J.~N. 2013, {FLAGCAL}: A flagging and calibration pipeline for
  {GMRT} data, Tech. rep., National Centre for Radio Astrophysics, Ganeshkhind,
  Pune

\bibitem[{{Cordes} \& {Lazio}(2002)}]{cl02}
{Cordes}, J.~M., \& {Lazio}, T.~J.~W. 2002, ArXiv Astrophysics e-prints,
  astro-ph/0207156

\bibitem[{{Crawford} {et~al.}(2001){Crawford}, {Gaensler}, {Kaspi},
  {Manchester}, {Camilo}, {Lyne}, \& {Pivovaroff}}]{cgk+01}
{Crawford}, F., {Gaensler}, B.~M., {Kaspi}, V.~M., {et~al.} 2001, \apj, 554,
  152

\bibitem[{{Cummings} \& {Campana}(2014)}]{cc14}
{Cummings}, J.~R., \& {Campana}, S. 2014, The Astronomer's Telegram, 6299, 1

\bibitem[{{Cummmings} {et~al.}(2014){Cummmings}, {Barthelmy}, {Chester}, \&
  {Page}}]{cbc+14}
{Cummmings}, J.~R., {Barthelmy}, S.~D., {Chester}, M.~M., \& {Page}, K.~L.
  2014, The Astronomer's Telegram, 6294, 1

\bibitem[{{Dodson} {et~al.}(2003){Dodson}, {Legge}, {Reynolds}, \&
  {McCulloch}}]{dlr+03}
{Dodson}, R., {Legge}, D., {Reynolds}, J.~E., \& {McCulloch}, P.~M. 2003, \apj,
  596, 1137

\bibitem[{{Duncan} \& {Thompson}(1992)}]{dt92}
{Duncan}, R.~C., \& {Thompson}, C. 1992, \apjl, 392, L9

\bibitem[{{Eatough} {et~al.}(2013){Eatough}, {Falcke}, {Karuppusamy}, {Lee},
  {Champion}, {Keane}, {Desvignes}, {Schnitzeler}, {Spitler}, {Kramer},
  {Klein}, {Bassa}, {Bower}, {Brunthaler}, {Cognard}, {Deller}, {Demorest},
  {Freire}, {Kraus}, {Lyne}, {Noutsos}, {Stappers}, \& {Wex}}]{efk+13}
{Eatough}, R.~P., {Falcke}, H., {Karuppusamy}, R., {et~al.} 2013, \nat, 501,
  391

\bibitem[{{Esposito} {et~al.}(2013){Esposito}, {Tiengo}, {Rea}, {Turolla},
  {Fenzi}, {Giuliani}, {Israel}, {Zane}, {Mereghetti}, {Possenti}, {Burgay},
  {Stella}, {G{\"o}tz}, {Perna}, {Mignani}, \& {Romano}}]{etr+13}
{Esposito}, P., {Tiengo}, A., {Rea}, N., {et~al.} 2013, \mnras, 429, 3123

\bibitem[{{Fich} {et~al.}(1989){Fich}, {Blitz}, \& {Stark}}]{fbs89}
{Fich}, M., {Blitz}, L., \& {Stark}, A.~A. 1989, \apj, 342, 272

\bibitem[{{Frail} {et~al.}(1999){Frail}, {Kulkarni}, \& {Bloom}}]{fkb99}
{Frail}, D.~A., {Kulkarni}, S.~R., \& {Bloom}, J.~S. 1999, \nat, 398, 127

\bibitem[{{Gold}(1969)}]{gol69}
{Gold}, T. 1969, \nat, 221, 25

\bibitem[{{Green}(2014)}]{gre14}
{Green}, D.~A. 2014, Bulletin of the Astronomical Society of India, 42, 47

\bibitem[{{He} {et~al.}(2013){He}, {Ng}, \& {Kaspi}}]{hnk13}
{He}, C., {Ng}, C.-Y., \& {Kaspi}, V.~M. 2013, \apj, 768, 64

\bibitem[{{Ibrahim} {et~al.}(2004){Ibrahim}, {Markwardt}, {Swank}, {Ransom},
  {Roberts}, {Kaspi}, {Woods}, {Safi-Harb}, {Balman}, {Parke}, {Kouveliotou},
  {Hurley}, \& {Cline}}]{ims+04}
{Ibrahim}, A.~I., {Markwardt}, C.~B., {Swank}, J.~H., {et~al.} 2004, \apjl,
  609, L21

\bibitem[{{Israel} {et~al.}(2014){Israel}, {Rea}, {Zelati}, {Esposito},
  {Burgay}, {Mereghetti}, {Possenti}, \& {Tiengo}}]{irz+14}
{Israel}, G.~L., {Rea}, N., {Zelati}, F.~C., {et~al.} 2014, The Astronomer's
  Telegram, 6370, 1

\bibitem[{{Israel} {et~al.}(2016){Israel}, {Esposito}, {Rea}, {Coti Zelati},
  {Tiengo}, {Campana}, {Mereghetti}, {Rodriguez Castillo}, {G{\"o}tz},
  {Burgay}, {Possenti}, {Zane}, {Turolla}, {Perna}, {Cannizzaro}, \&
  {Pons}}]{ier+16}
{Israel}, G.~L., {Esposito}, P., {Rea}, N., {et~al.} 2016, \mnras, 457, 3448

\bibitem[{{Kalberla} {et~al.}(2005){Kalberla}, {Burton}, {Hartmann}, {Arnal},
  {Bajaja}, {Morras}, \& {P{\"o}ppel}}]{kbh+05}
{Kalberla}, P.~M.~W., {Burton}, W.~B., {Hartmann}, D., {et~al.} 2005, \aap,
  440, 775

\bibitem[{{Leahy} \& {Tian}(2008)}]{lt08}
{Leahy}, D.~A., \& {Tian}, W.~W. 2008, \aap, 480, L25

\bibitem[{{Levin} {et~al.}(2010){Levin}, {Bailes}, {Bates}, {Bhat}, {Burgay},
  {Burke-Spolaor}, {D'Amico}, {Johnston}, {Keith}, {Kramer}, {Milia},
  {Possenti}, {Rea}, {Stappers}, \& {van Straten}}]{lbb+10}
{Levin}, L., {Bailes}, M., {Bates}, S., {et~al.} 2010, \apjl, 721, L33

\bibitem[{{Levin} {et~al.}(2012){Levin}, {Bailes}, {Bates}, {Bhat}, {Burgay},
  {Burke-Spolaor}, {D'Amico}, {Johnston}, {Keith}, {Kramer}, {Milia},
  {Possenti}, {Stappers}, \& {van Straten}}]{lbb+12}
{Levin}, L., {Bailes}, M., {Bates}, S.~D., {et~al.} 2012, \mnras, 422, 2489

\bibitem[{{Levine} {et~al.}(2006){Levine}, {Blitz}, \& {Heiles}}]{lbh06}
{Levine}, E.~S., {Blitz}, L., \& {Heiles}, C. 2006, Science, 312, 1773

\bibitem[{{Manchester} {et~al.}(2005){Manchester}, {Hobbs}, {Teoh}, \&
  {Hobbs}}]{mht+05}
{Manchester}, R.~N., {Hobbs}, G.~B., {Teoh}, A., \& {Hobbs}, M. 2005, \aj, 129,
  1993

\bibitem[{{Milne}(1968)}]{mil68}
{Milne}, D.~K. 1968, Australian Journal of Physics, 21, 201

\bibitem[{{Naidu} {et~al.}(2015){Naidu}, {Joshi}, {Manoharan}, \&
  {Krishnakumar}}]{njm+15}
{Naidu}, A., {Joshi}, B.~C., {Manoharan}, P.~K., \& {Krishnakumar}, M.~A. 2015,
  Experimental Astronomy, 39, 319

\bibitem[{{Olausen} \& {Kaspi}(2014)}]{ok14}
{Olausen}, S.~A., \& {Kaspi}, V.~M. 2014, \apjs, 212, 6

\bibitem[{{Pacini} \& {Salvati}(1973)}]{ps73}
{Pacini}, F., \& {Salvati}, M. 1973, \apj, 186, 249

\bibitem[{{Pavlovi{\'c}} {et~al.}(2013){Pavlovi{\'c}}, {Uro{\v s}evi{\'c}},
  {Vukoti{\'c}}, {Arbutina}, \& {G{\"o}ker}}]{puv+13}
{Pavlovi{\'c}}, M.~Z., {Uro{\v s}evi{\'c}}, D., {Vukoti{\'c}}, B., {Arbutina},
  B., \& {G{\"o}ker}, {\"U}.~D. 2013, \apjs, 204, 4

\bibitem[{{Rea} {et~al.}(2012){Rea}, {Pons}, {Torres}, \& {Turolla}}]{rpt+12}
{Rea}, N., {Pons}, J.~A., {Torres}, D.~F., \& {Turolla}, R. 2012, \apjl, 748,
  L12

\bibitem[{{Roy} {et~al.}(2010){Roy}, {Gupta}, {Pen}, {Peterson}, {Kudale}, \&
  {Kodilkar}}]{rgp+10}
{Roy}, J., {Gupta}, Y., {Pen}, U.-L., {et~al.} 2010, Experimental Astronomy,
  28, 25

\bibitem[{{Serylak} {et~al.}(2009){Serylak}, {Stappers}, {Weltevrede},
  {Kramer}, {Jessner}, {Lyne}, {Jordan}, {Lazaridis}, \& {Zensus}}]{ssw+09}
{Serylak}, M., {Stappers}, B.~W., {Weltevrede}, P., {et~al.} 2009, \mnras, 394,
  295

\bibitem[{{Stil} {et~al.}(2006){Stil}, {Taylor}, {Dickey}, {Kavars}, {Martin},
  {Rothwell}, {Boothroyd}, {Lockman}, \& {McClure-Griffiths}}]{std+06}
{Stil}, J.~M., {Taylor}, A.~R., {Dickey}, J.~M., {et~al.} 2006, \aj, 132, 1158

\bibitem[{{Str{\"u}der} {et~al.}(2001){Str{\"u}der}, {Briel}, {Dennerl},
  {Hartmann}, {Kendziorra}, {Meidinger}, {Pfeffermann}, {Reppin}, {Aschenbach},
  {Bornemann}, {Br{\"a}uninger}, {Burkert}, {Elender}, {Freyberg}, {Haberl},
  {Hartner}, {Heuschmann}, {Hippmann}, {Kastelic}, {Kemmer}, {Kettenring},
  {Kink}, {Krause}, {M{\"u}ller}, {Oppitz}, {Pietsch}, {Popp}, {Predehl},
  {Read}, {Stephan}, {St{\"o}tter}, {Tr{\"u}mper}, {Holl}, {Kemmer}, {Soltau},
  {St{\"o}tter}, {Weber}, {Weichert}, {von Zanthier}, {Carathanassis}, {Lutz},
  {Richter}, {Solc}, {B{\"o}ttcher}, {Kuster}, {Staubert}, {Abbey}, {Holland},
  {Turner}, {Balasini}, {Bignami}, {La Palombara}, {Villa}, {Buttler},
  {Gianini}, {Lain{\'e}}, {Lumb}, \& {Dhez}}]{sbd+01}
{Str{\"u}der}, L., {Briel}, U., {Dennerl}, K., {et~al.} 2001, \aap, 365, L18

\bibitem[{{Sun} {et~al.}(2011){Sun}, {Reich}, {Reich}, {Xiao}, {Gao}, \&
  {Han}}]{srr+11}
{Sun}, X.~H., {Reich}, P., {Reich}, W., {et~al.} 2011, \aap, 536, A83

\bibitem[{{Surnis} {et~al.}(2014){Surnis}, {Krishnakumar}, {Maan}, {Joshi}, \&
  {Manoharan}}]{skm+14}
{Surnis}, M.~P., {Krishnakumar}, M.~A., {Maan}, Y., {Joshi}, B.~C., \&
  {Manoharan}, P.~K. 2014, The Astronomer's Telegram, 6376, 1

\bibitem[{{Verschuur} {et~al.}(1974){Verschuur}, {Kellermann}, \& {van
  Brunt}}]{vkv74}
{Verschuur}, G.~L., {Kellermann}, K.~I., \& {van Brunt}, V. 1974, {Galactic and
  Extra-Galactic Radio Astronomy}, 127, doi:10.1007/978-3-642-96178-6

\bibitem[{{Younes} {et~al.}(2012){Younes}, {Kouveliotou}, {Kargaltsev},
  {Pavlov}, {G{\"o}{\v g}{\"u}{\c s}}, \& {Wachter}}]{ykk+12}
{Younes}, G., {Kouveliotou}, C., {Kargaltsev}, O., {et~al.} 2012, \apj, 757, 39

\bibitem[{{Yusef-Zadeh} {et~al.}(2015){Yusef-Zadeh}, {Diesing}, {Wardle},
  {Sjouwerman}, {Royster}, {Cotton}, {Roberts}, \& {Heinke}}]{ydw+15}
{Yusef-Zadeh}, F., {Diesing}, R., {Wardle}, M., {et~al.} 2015, \apjl, 811, L35

\end{thebibliography}

\end{document}